\documentstyle[11pt,aaspp4,tighten]{article}
\def\rosat{{\sl ROSAT~}}

\slugcomment{}
\begin{document}

\title{\bf X-ray Shapes of Distant  Clusters:\\
the Connection to Blue Galaxy Fractions}
\author{Q. Daniel Wang and Melville P. Ulmer}
\affil{Dearborn Observatory, Northwestern University}
\affil{ 2131 Sheridan Road, Evanston,~IL 60208-2900}
\affil{E-mail: wqd@nwu.edu and ulmer@ossenu.astro.nwu.edu}
\begin{abstract}

	Based on {\sl ROSAT} PSPC pointed observations, we have determined the 
aggregate X-ray shapes of 10 distant ($z = 0.17-0.54$) rich clusters: 
A2397, A222, A520, A1689, A223B, A1758, A2218, A2111, A2125, 
and CL0016+16. Four of the clusters have global X-ray ellipticities 
$\gtrsim 0.2$, as measured on a scale of diameter $\sim 3 h_{50}^{-1}$~Mpc.
These strongly elongated clusters tend to show substantial amounts of 
substructure, indicating that they are dynamically young systems.
Most interestingly, the global X-ray ellipticities of the clusters correlate 
well with their blue galaxy fractions; the correlation coefficient is 
0.75 with a 90\% confidence range of 0.44-0.92. This correlation suggests that 
blue cluster galaxies originate in the process of cluster formation, and that 
the blue galaxy proportion of a cluster decreases as the intracluster
medium relaxes onto equipotential surfaces.  

\end{abstract}
\keywords{cosmology: observations --- large-scale structure of universe ---
galaxies: clusters: general --- galaxies: evolution --- X-rays: general}

\section {Introduction}

	By comparing the rest-frame colors of cluster galaxies relative to 
those of early-type galaxies at the same epoch, Butcher \& Oemler (1984; 
BO84 hereafter) found that the blue galaxy fraction $f_b$ in rich clusters 
decreases rapidly with time since $z \sim 0.4$. This so-called Butcher \& 
Oemler effect has also been confirmed spectroscopically 
(Dressler \& Gunn 1992 and references therein). While the true nature of the 
effect remains uncertain (Oemler et al. 1997 and references therein),
finding a connection of the blue galaxy proportion to other cluster properties
may provide insights into the origin and evolution of blue cluster galaxies.

 	We here report the detection of a correlation between the $f_b$ 
values and the global X-ray ellipticities of 10
distant ($z \gtrsim 0.1$) clusters that were optically 
surveyed by BO84. We first discuss the selection of the clusters, the X-ray
observations, and the data reduction (\S 2), and we then briefly describe 
both the algorithm used for the X-ray morphological analysis and Monte-Carlo
simulations to assess systematic effects (\S 3). We present the analysis of 
the correlation in \S 4. Since only the projected shapes of the 
clusters are measured here, we start \S 5 by acknowledging ambiguities 
caused by projection effects, and we then consider both 
the origin and evolution of cluster elongation
and how they relate to the Butcher \& Oemler effect. Finally in \S 6,
we summarize our results and conclusions. As in BO84 we adopt 
$H_o = 50 {\rm~km~s^{-1}~Mpc^{-1}}$ and $q_o = 0.1$ throughout the paper.

\section{X-ray Images}

	Table 1 lists the salient parameters of both our selected clusters 
and \rosat PSPC observations (T\"umper 1992 and references therein). 
As a measure of cluster optical richness, 
the parameter $N_{30\%}$ is between 23 and 155 for these clusters (Table 1), 
compared to 21 and 94 for Virgo and Coma --- two best-known nearby clusters. 
The only other distant BO84 cluster
that was also observed with the PSPC, but is not included in this study, is
A777 ($z=0.226$, $f_b =  0.05\pm0.08$, $N_{30\%}=15$). This cluster is the 
poorest in the BO84 sample, and its PSPC observation shows little (if any) 
diffuse emission from the cluster. The observations of A2111 and A520 were 
obtained by us, while the other seven (both A222 and A223 were covered by
the same observation) were retrieved from the \rosat archive
(http://heasarc.gsfc.nasa.gov/). We constructed X-ray images 
in the 0.5-2~keV band (PSPC channels: 52-201; Fig. 1) to maximize 
cluster-to-background contrasts (Wang, Ulmer, \& Lavery 1997). 

\begin{table}[htb] 
\scriptsize
\begin{tabular}{lcrrcccccc}
\multicolumn{10}{c}{\bf Table 1}\\
\multicolumn{10}{c}{\bf Cluster Parameters and X-ray Observations\tablenotemark{a}}\\[0.1in]
\hline \hline                      
Cluster& $z$  & $f_b$&$N_{30\%}$  &\rosat & Exposure &\multicolumn{2}{c}{ellipticity}& \multicolumn{2}{c}{Position Angle (deg, N-E)} \\
	& 	&(\%)& 		  &No.		&(s) & $D=3$~Mpc &$D=1.5$~Mpc&$D=3$~Mpc&$D=1.5$~Mpc\\
\hline	
A2397  & 0.222& $ 1\pm3$&  23& wp800344& 13152&0.07(0.00-0.15)&0.08(0.00. 0.19)&155(117-213)&230(161-276)  \\
CL0016+16&0.541&$ 2\pm7$&  65& rp800253& 40562&0.07(0.00-0.11)&0.21(0.15-0.27)&61(36-84)    &50(42-57)   \\
A222   & 0.211& $ 6\pm4$&  45& rp800048&  6780&0.10(0.02-0.26)&0.27(0.17-0.45)&70(41-129)   &98(75-113)  \\
A520   & 0.203& $ 7\pm7$& 126& rp800480&  4592&0.08(0.00-0.15)&0.17(0.06-0.25)&96(27-159)   &11(0.8-31)  \\
A1689  & 0.175& $ 9\pm3$& 124& rp800248& 13957&0.16(0.12-0.19)&0.14(0.11-0.17)&183(176-189 )&22(16-27)   \\
A1758  & 0.280& $ 9\pm4$&  91& rp800047& 13509&0.16(0.10-0.23)&0.42(0.33-0.46)&152(139-169) &123(119-129)\\
A223B  & 0.207& $10\pm6$&  67& rp800048&  6780&0.53(0.45-0.58)&0.23(0.10-0.33)&22(17-26)    &183(164-212) \\
A2218  & 0.171& $11\pm4$& 114& wp800097& 37014&0.24(0.22-0.27)&0.20(0.18-0.23)&96(92-100)   &91(86-95)   \\
A2111  & 0.228& $16\pm3$& 155& rp800479&  7288&0.36(0.27-0.46)&0.39(0.33-0.46)&153(145-160) &145(140-152)\\
A2125  & 0.247& $19\pm3$&  62& rp800511& 18352&0.38(0.29-0.48)&0.37(0.27-0.44)&132(124-140) &127(118-135)\\
\hline 
\end{tabular}
\tablenotetext{a}{The redshift $z$, the blue galaxy fraction $f_b$, and 
the 30\% of the total number of cluster members (brighter than $M_v \sim -20$)
$N_{30\%}$ are from Butcher \& Oemler (1984). Uncertainties in parameter 
values, as presented in parentheses, are at the 90\% confidence level.
}
\end{table}

	We excised from the X-ray images point-like sources detected
with signal-to-noise ratios $\gtrsim 3$. The PSPC count distributions of
these sources are consistent with the instrument point response function (PSF;
$\sim 25\arcsec$ FWHM on-axis) at a confidence $\gtrsim 5\%$.  
The data within the 90\% source flux-encircled radius around each source were 
replaced by randomly generated events of the intensity interpolated from
neighboring pixels. From Monte Carlo simulations (\S 3), we find that
residual source fluxes in the images produce negligible effects on
our measurement of global X-ray shapes of the clusters, in comparison to
statistical uncertainties. 

	A few of the images are contaminated by neighboring
diffuse X-ray objects. A1758: This cluster has a companion about 8\arcmin\ 
south (the distance between the cluster centroids; Mushotzky 1992). A small 
portion of this companion, whose redshift 
is yet unknown, appears in the bottom of the A1758 image. 
A223B: This cluster image also contains A223A ($z=0.206$; 
Sandage, Kristian, \& Westphal 1976) and a small section 
of A222 at the lower right corner. The redshift differences, 
together with the large projected separations in the sky, indicate that 
these clusters are not physically interacting with each other.
A223A, about 4\arcmin\ northeast to A223B, is much fainter in X-rays than 
A223B. A2125: This cluster is associated with a filamentary 
feature of low surface brightness diffuse X-ray emission 
(Wang, Connolly, Brunner 1997), part of which can be seen at the southwestern
corner of the image. In order to avoid these extraneous features, 
we chose a maximum scale of {\sl diameter} $D \sim 3$~Mpc to characterize 
the global X-ray shapes of the clusters. 

\section{X-ray Shapes}

	We characterized the aggregate X-ray shapes of the clusters, 
using ellipse parameters (centroid, ellipticity $\epsilon$, and 
position angle). An iterative algorithm for the parameter computation
has been detailed by Wang, Ulmer, \& Lavery (1997).
Here we only outline the procedure. For each cluster, we started an iteration
with the calculation of the first and second moments of the source-excised 
PSPC count distribution within a circular region of diameter $D$ around an 
assumed cluster centroid. From the moments, we then derived
the ellipse parameters, which defined an elliptical region of major axis
$D$ for the next run. This iteration went on until all the ellipse parameters 
converged with relative parameter changes all less than  $10^{-3}$. 
In each iteration we subtracted background contributions to the moments.
The background level was estimated in an annulus between
10\arcmin-16\arcmin\ radii from the cluster centroid. We estimated 
the statistical uncertainties in the final ellipse parameters, using 1000 
replications from bootstrapping realizations of the count distribution.
vThe 90\% confidence interval of each parameter, for example, represents
the 5\% and 95\% percentiles of the replicated parameter values. 
Table 1 includes the ellipse parameters measured at two representative
scales of $D = 1.5$~Mpc and 3~Mpc.
 
	We also conducted extensive Monte-Carlo simulations to assess 
systematic uncertainties. Simulated images contain about the same cluster 
and background counts as observed, and approximate the cluster morphologies
with the elliptical $\beta$ model (e.g., Neumann \& B\"ohringer 1996)
of various ellipticities. We randomly added point-like X-ray sources in the 
simulated images, according to the source luminosity function given by 
Hasinger et al. (1994). Bright sources were detected and excised as in real
X-ray images (\S 2).

	We then applied the above algorithm to these simulated images, 
which typically reproduced the model parameters well. But for models of 
$\epsilon \lesssim 0.05$, the algorithm tended to overestimate the 
ellipticities. This problem is also present in previous studies of
cluster shapes, using similar algorithms (e.g., Carter \& Metcalfe 
1980; McMillan,  Kowalski, \& Ulmer 1989; Buote \& Canizares 1996). 
A correction for the problem, which we have not attempted,
would only slightly enhance the $\epsilon-f_b$ correlation to be discussed 
in \S 4. Furthermore, we find that the effect of residual point-like
sources in the X-ray images is negligible in comparison to
statistical errors. A superposition of a cluster with an extended 
source could, however, alter the X-ray morphology of the cluster 
substantially. But such superpositions with the cores ($D \lesssim 1.5$~Mpc)
should be rare occasions for these distant clusters 
($\sim 10^{-1}$). A superposition in regions
away from cluster cores can easily be recognized, e.g., the
A223A/A223B pair. Therefore, these systematic uncertainties 
should not be important in our study of the {\sl global} X-ray shapes of the 
clusters. 

\section {The $\epsilon-f_b$ Correlation}

	Fig. 2 shows an apparent correlation between $\epsilon$ and $f_b$. 
Except for A223B, the figure uses the ellipticities as measured on a scale of 
$D \sim 3$~Mpc (Table 1).  Around this scale,
the ellipticities typically show little change and is not 
sensitive to subcomponents occasionally seen in core regions. 
But for A223B the figure uses the ellipticity 
measured at $D \sim 1.5$~Mpc to avoid A223A, which is projected at a
distance of $\sim 1.1$~Mpc from A223B (\S 2.2). We resorted 
to the bootstrapping method again to quantify the correlation. For 
each of the 1000 bootstrapping replications of $\epsilon$ 
(\S 3), we randomly generated a value of $f_b$, assuming a Normal distribution 
with the best estimate and 1$\sigma$ dispersion as listed in Table 1. 
We made all negative $f_b$ values be zero. Although a measured $f_b$ could 
be negative due to the statistical uncertainty in the background galaxy 
subtraction, a negative $f_b$ is physically not real. 
We calculated the Pearson correlation coefficient for each set of the 
replications, and then sorted the coefficients from small to large. 
The median and the 5\% and 95\% percentiles of the sorted coefficients 
gave the best estimate and the 90\% confidence limits of the 
coefficient as 0.75(0.44-0.92). 

	The correlation is less significant on smaller scales. 
For example, the correlation coefficient is 0.51(0.15-0.77) on the 
scale of $D \sim 1.5$~Mpc. This weaker correlation is due to 
the presence of subcomponents in cluster cores. These small-scale 
subcomponents show little correlation with the global aggregate shapes of
the clusters.

\section{Discussion}

\subsection{Projection Effects}

	While the observed X-ray surface brightness is sensitive only to 
the emission measure of the hot intracluster medium (ICM), 
we cannot disentangle the projection effect for individual 
clusters with the X-ray images alone. Thus, a circular morphology of a cluster 
does not necessarily mean a spherical symmetric distribution of the ICM. 
Projection effects cannot, however, produce an elongated shape of an 
intrinsically round cluster. The net result of the projection effects
is a statistically weaker $\epsilon - f_b$ correlation, and the intrinsic 
correlation could actually be stronger than the observed.
The projection effects should be small, if strongly elongated clusters tend to
have triaxial and/or irregular ICM morphologies. Since we are primarily 
interested in the statistical properties of the clusters, we neglect 
projection effects in our qualitative discussion below. 

\subsection{Origin of the Elongated X-ray Morphologies}

A2111 and A2125, the two clusters with the largest ellipticities in Fig. 2,
are clearly dynamically young systems. A2111 also shows the greatest centroid 
shift with scale in our cluster sample. A detailed 
optical and X-ray study of the cluster (Wang, Ulmer, \& Lavery 1997)
suggests that this cluster is an ongoing coalescence of at least two 
subclusters. A2125 has a multiple-peak morphology and is associated with 
a hierarchical superstructure of galaxies, clusters and diffuse hot gas,
indicating that the cluster is still at its early stage of formation
(Wang, Connolly, Brunner 1997). Therefore, the elongated X-ray morphologies 
are likely produced during the formation of the clusters.

	Theoretically, cluster formation vie gravitational attraction naturally
results in an elongated mass distribution. Relevant processes include 
the tidal 
distortion on the proto-cluster by neighboring objects (Binney \& Silk 1979), 
the triaxial collapse of the density perturbation (Elsenstein \& Loeb 1995), 
and the merger between subunits (e.g., Evrard et al. 1993). 
When a cluster is just formed, the ICM, following the {\sl mass} distribution
of the dark matter, can thus have a strongly elongated and lumpy morphology, 
which is manifested by the X-ray emission.

\subsection {Evolution of the X-ray Morphologies}

	As the isotropic pressure tensor of the ICM  drives
it to lie on equipotential surfaces of a cluster, the aggregate X-ray 
shape evolves to become smoother and rounder (e.g., 
Evrard et al. 1993). Hence we argue here that the global X-ray roundness is
a cluster age indicator. Nearby rich clusters, more-or-less relaxed,
do show significantly rounder X-ray morphologies than the mass
distributions (Buote \& Canizares 1996).
The relaxation should happen on a few sound crossing times in 
the ICM. The crossing time can be estimated as $t_s \sim (4 \times 10^9 
{\rm~years}) (T_g/3 \times 10^7 {\rm~K})^{-1/2} (D/3 {\rm~Mpc})$, where 
the ICM temperature $T_g$ may fall considerably from the
 core to the outskirts of a cluster, and the scaled value is the characteristic mean
temperature of the high $\epsilon$ clusters 
(A2111 --- Wang, Ulmer, \& Lavery 1997; A2125 --- Wang, Connolly, \&
Brunner 1997). In comparison, the timescale for the cluster to wipe out 
a subcomponent of a few times hundred kpc across is  
$\sim 10^9$~years. Such subcomponents presumably represent 
merging subgroups, whose X-ray emission can be strongly enhanced by
the high pressure in cluster cores. Numerical simulations of hierarchical
structure formation do show that the global X-ray shape of a cluster
tends to become increasingly round with time (e.g., Evrard et al. 
1993; Katz \& White 1993; Buote \& Tsai 1995). Therefore, the roundness of 
the global X-ray shape is a measure of the relaxation state of a cluster.

	Since the relaxation timescale for the global X-ray shape of a 
cluster can be comparable to the epoch duration $\sim 9 \times 10^{9} 
{\rm~years}$ from  $z \sim 0.4$ to $z \sim 0.1$, one might expect to see 
statistical evidence for the cluster X-ray shape evolution with $z$. We, 
therefore, compare our results with the similar X-ray study of five nearby 
($< 0.1$) clusters (Buote \& Canizares 1996): A401, A1656 (Coma), A2029, 
A2199, and A2256. The optical richness of these clusters is comparable
to those in Table 1. Buote \& Canizares presented 
in their Table 4 the aggregate X-ray 
ellipticities measured on various scales for each of the five clusters,
We estimate $\epsilon \sim 0.17, 0.14,$ and 0.22 on a scale of 
$D \sim 3$~Mpc for A401, A2029, A2256. For A1656 and A2199, 
$\epsilon \sim 0.20$ and 0.14 on the largest scales presented
$D \sim 1.9$~Mpc and 2.6~Mpc, respectively; the ellipticities at $D \sim 3$~Mpc
should be smaller. The errors in these measurements 
are typically $\lesssim 15\%$. The relatively 
large ellipticity of A2256 is due to an exceptionally prominent 
subcomponent of the cluster (e.g., Briel \& Henry 1994). 
In comparison, four out of the ten distant clusters in Table 1 have 
$\epsilon \gtrsim 0.2$ at $D \sim 3$~Mpc. Thus the evidence 
for the evolution, if present, is still weak, which may be due to 
the small number statistics presented here and to the large dispersion in 
cluster ages at every epoch. 

\subsection{Implications for the Butcher \& Oemler Effect}

In contrast, the correlation between the blue galaxy fraction and the global 
X-ray shape is strong enough to show up in a sample of 10 objects, indicating
an intrinsic connection between galaxy evolution and cluster dynamics.
This connection is consistent with the notion the Butcher \& Oemler effect 
is due to the transformation of 
late-type galaxies into early-type ones in clusters formed through 
hierarchical clustering (Kauffman 1995; Oemler et al. 1997). In a hierarchical 
clustering scenario, a typical distant cluster is assembled from {\sl smaller} 
units and over a {\sl shorter} period than a nearby cluster of the same mass. 
Because smaller units tend to contain more gas-rich late-type
galaxies, distant clusters are {\sl born} with larger proportions of such 
galaxies than nearby ones. Blue galaxies observed in distant clusters are 
indeed typically gas-rich, starforming, or even starburst spirals/irregulars 
(e.g., Lavery \& Henry 1994; Oemler et al. 1997). 
Starbursts can be triggered by gas compression, due to both the high ICM 
pressure and the tidal forces from the mean cluster potential and 
from frequent encountering cluster members (Henriksen \& Byrd 1996; 
Moore et al. 1996). 
Such ``galaxy harassment'' by the cluster environment can also transform
late-type blue galaxies to early-type red ones, resulting in the decline
of the blue cluster galaxy proportion with time (Oemler et al. 1997).
While details about this transformation are still elusive, 
the presence of the $\epsilon-f_b$ correlation suggests that the 
transformation process operates on a timescale similar to that of the ICM 
relaxation from a strongly elongated morphology at the cluster's birth to
a more-or-less round shape. 

\section {Summary and Conclusion}

	We have measured the aggregate X-ray shapes of 10 BO84 clusters
in the redshift range between $z = 0.17-0.54$. Strongly elongated clusters 
typically show irregular X-ray morphologies (e.g., multiple peaks and 
centroid shifts), and are probably
formed only recently as the coalescences of subunits. 
Because the ICM tends to relax onto the equipotential surfaces of a cluster,
the roundness of the global X-ray shape is a
good indicator of the dynamic relaxation state of the ICM.

	The  ellipticity measured at diameter $D \sim 3$~Mpc of a
cluster correlates well with the blue galaxy fraction.
This correlation results naturally from the combined action of
the cluster formation through hierarchical clustering and the environmental 
transformation of cluster galaxies from late-types into early-types,
if the transformation takes place on a timescale comparable to that of the 
global ICM relaxation.

	This connection of the global X-ray shape of a cluster to both the
ICM relaxation state and the blue galaxy proportion, if quatified with
improved number statistics, will offer insights into environmental
effects on cluster galaxy evolution and will provide a powerful tool for 
studying distant clusters.

\acknowledgements

	This project is supported by NASA under Grant 5-2717.

\twocolumn

\begin{figure}
\figurenum{1}
\epsscale{0.65}
\plotone{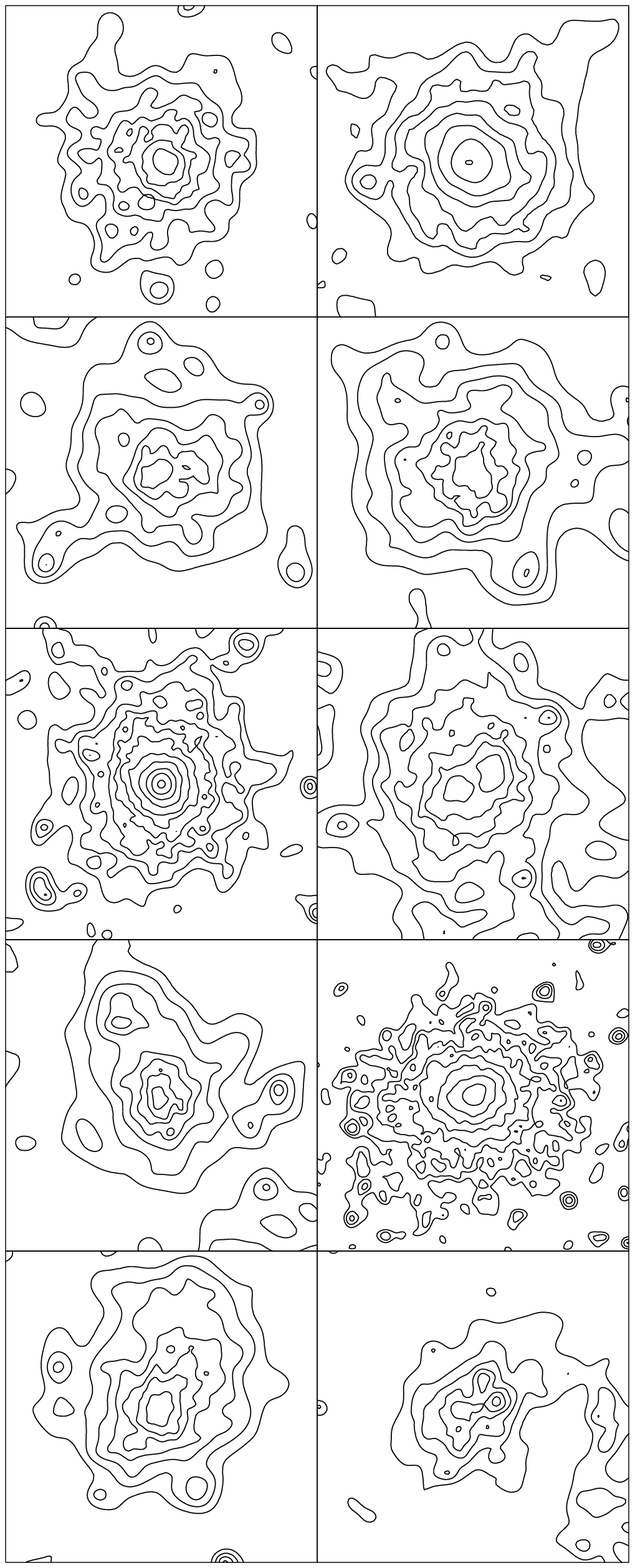}
\caption{ 0.5-2~keV band PSPC images of the clusters:
A2397, CL0016+16, A222, A520, A1689, A1758, A223B, A2218, A2111, and 
A2125 (left to right and top to bottom), in order of the $f_b$ values 
(Table 1). The images all have the same physical 
scale of 4~Mpc on a side, and have their angular scales of 14.1, 8.2, 14.6, 
15.2, 16.8, 12.0, 14.8, 17.1, 13.8 and 13.0 arcmin, respectively.
The images, exposure-corrected and background-subtracted and source-excised, 
are smoothed adaptively with a Gaussian, the size of which is adjusted at 
each pixel to achieve a count-to-noise ratio of 4. Each contour is a factor
of two of its lower level; the lowest contour is at $2.5 \times 10^{-4} 
{\rm~counts~s^{-1}~arcmin^{-2}}$. 
\label{fig1}}
\end{figure}

\begin{figure} 
\figurenum{2}
\epsscale{1}
\plotone{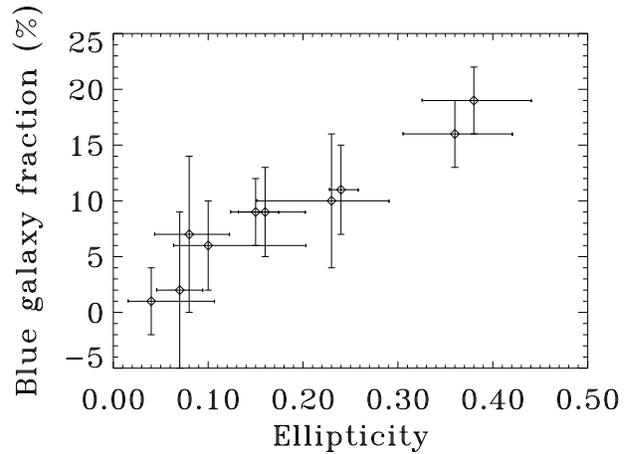}
\caption{Blue galaxy fractions vs. global X-ray ellipticities 
as measured on a scale of major axis equal to 3~Mpc.
The error bars are all at the 1$\sigma$ level.
\label{fig2}}
\end{figure}

\end{document}